\documentstyle[12pt,aaspp,psfig]{article}
\newcommand{\Ab}{A_{\rm b}}
\newcommand{\Bb}{B_{\rm b}}
\newcommand{\Ao}{A_{\rm o}}
\newcommand{\Bo}{B_{\rm o}}
\newcommand{\Ap}{A_{\rm p}}
\newcommand{\Bp}{B_{\rm p}}
\newcommand{\cntr}[1]{\multicolumn{1}{c}{#1}}
\begin{document}

\title{The detection and photometric redshift determination of distant
galaxies using {\it SIRTF\/}'s Infrared Array Camera}

\author{Chris Simpson\altaffilmark{1} and Peter Eisenhardt}
\affil{Jet Propulsion Laboratory, California Institute of Technology, Mail
Stop 169--327, 4800 Oak Grove Drive, Pasadena, CA 91109}

\altaffiltext{1}{Present address: Subaru Telescope, National
Astronomical Observatory of Japan, 650 N. A`oh\={o}k\={u} Place, Hilo,
HI 96720}

\begin{abstract}
We investigate the ability of the {\it Space Infrared Telescope
Facility\/}'s Infrared Array Camera to detect distant ($z \sim 3$)
galaxies and measure their photometric redshifts. Our analysis shows that
changing the original long wavelength filter specifications provides
significant improvements in performance in this and other areas.
\end{abstract}
\keywords{
cosmology: early universe ---
galaxies: evolution ---
infrared: galaxies ---
instrumentation: photometers ---
techniques: photometric
}

\section{Introduction}

On 1998 March 25, NASA Administrator Daniel Goldin approved the
initiation of final design and construction for the {\it Space
Infrared Telescope Facility\/} ({\it SIRTF\/}; Fanson et al.\ 1998).
With launch planned for December 2001, {\it SIRTF\/} will complete
NASA's family of Great Observatories. {\it SIRTF\/} will provide a
large increase in sensitivity over previous missions across its
3--200\,\micron\ operating range, and over 75\% of the observing time
during its 2.5\,year-minimum lifetime will be awarded to general
investigators. A call for Legacy Proposals (large projects of both
immediate scientific interest and lasting archival value, and with no
proprietary data period) is planned for July of 2000.

One of the defining scientific programs for {\it SIRTF\/} is the study
of galaxies to $z > 3$ by means of deep surveys at 3--10\,\micron.
This limit was selected because it is apparently beyond the peak in
the space density of luminous quasars (Schmidt, Schneider, \& Gunn
1995).  Not only will {\it SIRTF\/}'s excellent sensitivity in this
wavelength region allow such galaxies to be detected with the Infrared
Array Camera (IRAC; Fazio et al.\ 1998), but the H$^-$ opacity minimum
at 1.6\,\micron\ (John 1988) is expected to be a major tool in
photometric redshift determination at $1 \lesssim z \lesssim 5$
(Wright, Eisenhardt \& Fazio 1994), since it is a ubiquitous feature
of stellar atmospheres.

UV-bright examples of such galaxies have already been detected by
means of the ``UV dropout'' technique (Steidel \& Hamilton 1992, 1993;
Steidel et al.\ 1996) and they play an important role in the overall
star formation history of the Universe (Madau et al.\ 1996). By
detecting galaxies on the strength of their UV emission, however, UV
dropout samples are necessarily biased in favor of those with both
active star formation and relatively modest extinction.  Such samples
will not reveal if there is an underlying population of galaxies which
have already assembled the bulk of their stellar mass.  In the absence
of ongoing star formation, even massive galaxies will be too faint in
the rest-frame ultraviolet to be picked up by optical surveys.  The
stellar mass already present at an early epoch constrains the star
formation rate to that point, a quantity which is still uncertain due
to the possibility of significant dust extinction. Steidel et al.\
(1998) suggest that ``the onset of substantial star formation activity
in galaxies occurs at $z > 4.5$'', and estimate the net extinction
correction to the star formation rate derived from UV dropouts to be a
factor of about 5. Recent far-infrared detections of the cosmic
background (Hauser et al.\ 1998), and far-infrared and submillimeter
detections of field galaxies with ISO and SCUBA (Kawara et al.\ 1998;
Puget et al.\ 1999; Barger et al.\ 1998; Hughes et al.\ 1998) also
hint at a heavily dust enshrouded starburst population. Hence an
accurate picture of the star formation history of the universe can
only be determined by making an accurate census of {\em all\/}
galaxies, not just star-forming ones with low extinction. Since the
luminosity in the rest-frame near-infrared correlates linearly with
mass (Gavazzi, Peirini \& Boselli 1996) and is relatively unaffected
by dust obscuration, this is clearly the spectral region in which to
make such a census.

In this paper we examine changing the specifications of the IRAC
filters in order to optimize them for the study of $z \sim 3$ galaxies,
including photometric redshift determination. IRAC will have two
detector arrays made from indium antimonide (InSb) which is sensitive
to $\lambda \lesssim 5$\,\micron\ photons, and two made from
arsenic-doped silicon (Si:As) which detects longer-wavelength light.
All four arrays will have a $256 \times 256$ format, with 1\farcs2
pixels, and will operate simultaneously. The InSb arrays will have
filters centered at 3.63\,\micron\ with 20.4\% bandwidth and at
4.53\,\micron\ with 23.3\% bandwidth. We shall refer to these filters
as $L$ and $M$, respectively, because of their similarity to the
filters used in ground-based facilities. The filters for the Si:As
arrays have no ground-based analogs, since they cover wavelengths where
the atmosphere is opaque, and so we refer to them as $A$ and $B$, with
$A$ being the bluer of the two. The $L$ and $A$ filters share one field
and are separated by a dichroic filter, while the $M$ and $B$ filters
similarly share a second nearly adjacent field. The original, or
``baseline'', specifications for $A$ and $B$ are 25\% bandwidth filters
centered at 6.3 and 8.0\,\micron.  Our analysis will involve comparing
the performance of different pairs of filters, and we distinguish
between these pairs by the use of subscripts, so for example the
baseline filters described above are called $\Ab$ and $\Bb$. We also
introduce the notation $[\lambda_1:\lambda_2]$ to describe a filter
which extends from $\lambda_1$\,\micron\ to $\lambda_2$\,\micron. In
this notation, the $L$ filter is $[3.26:4.00]$, $M = [4.00:5.06]$, $\Ab
= [5.51:7.09]$, and $\Bb = [7.00:9.00]$.

\section{Assumptions}

The calculations in this paper all assume a total integration time of
10\,hours per point (made up of many individual 200\,second
exposures), which allows a deep image of a large area of sky to be
obtained in a sensible integration time. For such exposure times, IRAC
requires background-limited operation and therefore we neglect read
noise and dark current. Confusion noise will be discussed in \S5.1 but
is otherwise not considered here. As {\it SIRTF\/} is required to be
diffraction limited at 6.5\,\micron, we adopt a detection area of
twenty pixels, independent of wavelength, and assume that all the
source flux is contained within this region.

\subsection{Telescope and instrument efficiency}

The fraction of photons incident upon {\it SIRTF\/}'s 85\,cm diameter
aperture which reach the IRAC detectors is specified in the {\it
SIRTF\/} Observatory Performance and Interface Control Document and
the IRAC Instrument Performance Requirements Document\footnote{The
{\it SIRTF\/} PICD is Jet Propulsion Laboratory document
674--SEIT--100, V2.2. The IPRD is Smithsonian Astrophysical
Observatory document IRAC96--202, revision 4.2.}, and is about 50\%,
including filter transmission. The InSb quantum efficiency (QE) is
assumed to be 80\% over the entire wavelength range of the two InSb
filters. The QE of the Si:As arrays is based on the average of
laboratory measurements made on a number of flight-candidate arrays
coated with anti-reflection coatings of either SiO or ZnS. Due to the
high opacity of SiO beyond 8\,\micron, in the following analysis we
assume that the ZnS coating will be used in preference to the SiO
coating if the filter bandpass extends beyond 8\,\micron. The net
result of these assumptions is a QE of approximately 60--65\% for the
two Si:As filters. We also assume that the various filters considered
here are opaque outside their nominal wavelength region.

\subsection{Background}

From the PICD, the minimum {\it SIRTF\/} background (which occurs at the
ecliptic pole for wavelengths considered here), $b(\lambda$) is computed
as
\[
b(\lambda) = 5.5 \times 10^{-14} B_\lambda(5500) + 4.2 \times 10^{-8}
B_\lambda(278.5) + 2.5 \times 10^{-4} (\lambda / 100\,\mu{\rm m})^2
B_\lambda (17.5) + B_\lambda (2.73) ,
\]
where $B_\lambda (T)$ is the Planck function for a temperature of $T$\,K.
The first term is due to scattered sunlight from zodiacal dust, the second
and third terms to thermal emission from zodiacal and Galactic dust,
respectively, and the final term is the cosmic microwave background
radiation. This function provides an excellent fit to data from the DIRBE
instrument (Hauser 1996).

\subsection{Cosmology and galaxy models}

Our analysis is concerned with the detection of high redshift galaxies.
We construct model spectra of such galaxies using the GISSEL96 spectral
synthesis code (Bruzual \& Charlot 1993, 1998; see also Charlot, Worthey
\& Bressan 1996). This code does not include polycyclic aromatic
hydrocarbon (PAH) features at 3.3\,\micron\ and beyond, but at $z = 3$
these features are at observed wavelengths beyond the range considered
here. We will return to the PAH features when we discuss confusion noise.

\begin{figure}
\plotfiddle{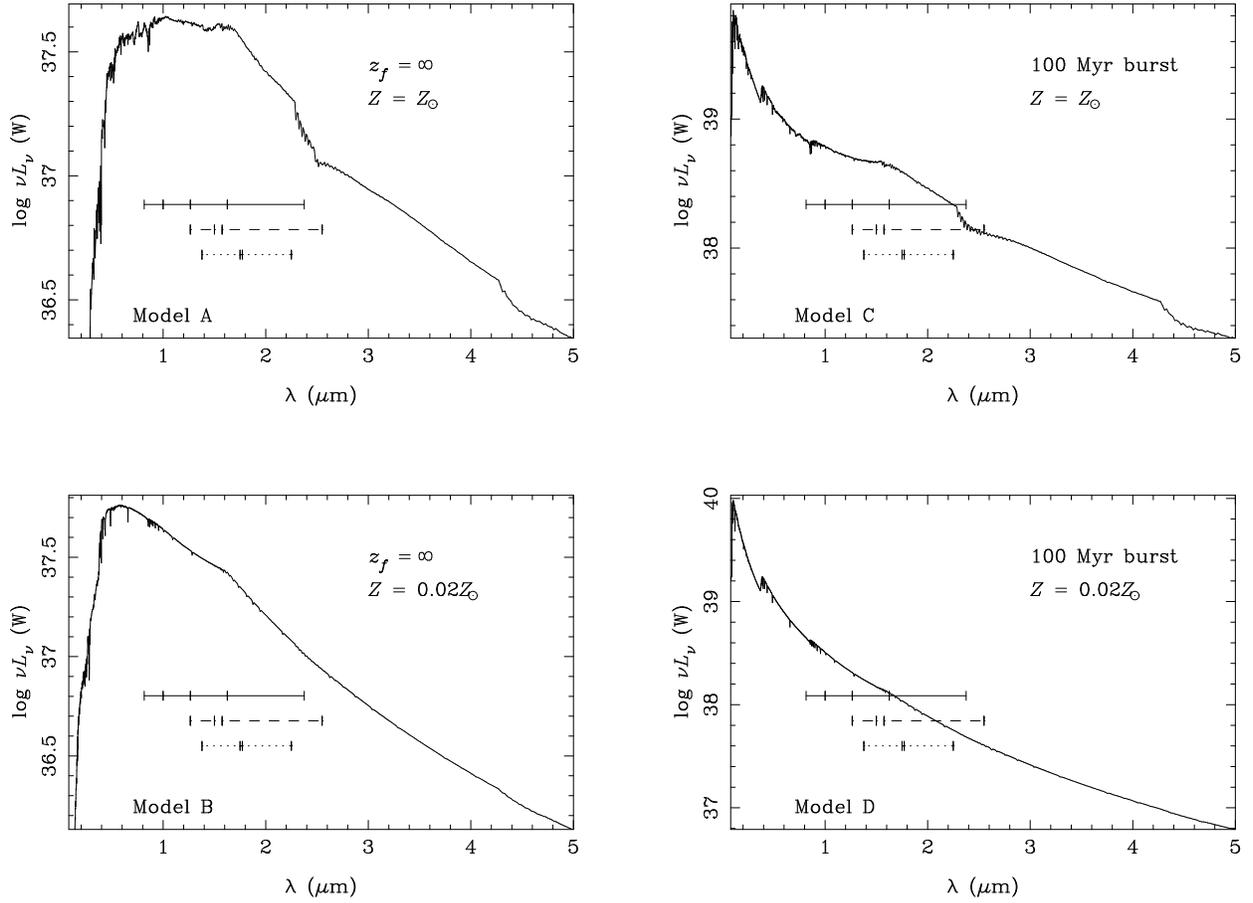}{360pt}{-90}{66}{66}{-250}{380}
\caption[]{The rest-frame spectra at $z = 3$ of the four galaxy models.  
The solid horizontal bars indicate the preferred filter set, while the
dashed bars show the optimum Si:As filters and the dotted bars the
baseline Si:As filters.}
\label{fig:spectra}
\end{figure}

We adopt a conservative (in the sense of making $z \sim 3$ galaxies
faint) cosmology with $H_0 = 50$\,km\,s$^{-1}$\,Mpc$^{-1}$ and $q_0 =
0.1$, corresponding to a present age of the Universe of 16.5\,Gyr (at $z
= 3$ the Universe is 2.9 Gyr old). We assume a Salpeter (1955) IMF
extending from 0.1--125\,$M_\odot$, and normalize our galaxies to have
$M_K = -24.63$ ($L^*$; Gardner et al.\ 1997) at an age of 16.5\,Gyr. We
use four models to represent the limits of galaxy evolution. Model A is a
maximally old model with solar metallicity formed in an instantaneous
burst at $z = \infty$ and evolving passively thereafter. Model B is like
A but with metallicity only $\frac{1}{50}$ solar. As we show in
\S3.3, these types of galaxies will not appear in  UV dropout samples.
Model C is a young solar metallicity galaxy formed with a constant
star formation rate over the 100\,Myr prior to whatever redshift is
considered (i.e.\ a young nonevolving spectrum).  Model D is like C,
but again with $Z = 0.02 Z_\odot$. Models C and D are much more
luminous than models A and B at $z = 3$, and are representative of the
spectra expected for UV-dropout galaxies. The rest frame
ultraviolet--optical--infrared spectra of these four models at $z = 3$
are shown in Figure~\ref{fig:spectra}.

The only feature common to all these spectra is a change in slope at
1.6\,\micron, due to the H$^-$ opacity minimum, although this is
barely discernable in Model D. It is in fact the only feature in the
two low-metallicity models, which lack the prominent CO bandhead at
2.3\,\micron\ and the effects of metal line blanketing at rest-frame
optical wavelengths. This therefore makes it a suitable feature for
use in photometric redshift determination, a technique whereby an
approximate redshift can be inferred from observed colors as one or
more spectral features pass through broad-band filters (e.g., Hogg et
al.\ 1998).

\section{Sensitivity}

Since {\it SIRTF\/}'s observations will be dominated by general
observer programs, it is important to consider performance independent
of any particular program, in addition to the high redshift galaxy
observations which drive the filter recommendations made here. In
fact, as we discuss in \S6, the pair of filters which provide the
optimum performance for $A-B$ color selection and photometric redshift
determination at $z \sim 3$ are not very suitable for other science
programs. For this reason we will throughout the course of this paper
be discussing the relative performance of three different Si:As filter
pairs, which we designate ``baseline'', ``optimum'', and ``preferred''
and denote by subscripts $b$, $o$, and $p$, respectively. The baseline
filters are the two 25\% bandwidth filters centered at 6.3 and
8.0\,\micron, the optimum filters provide the most accurate $A-B$
color selection and photometric redshift determination (see
\S4) at $z \sim 3$, and the preferred filters give slightly poorer
performance in these areas, but are expected to be much more useful
for other scientific programs. The optimum filters are $\Ao =
[5.06:6.00]$ and $\Bo = [6.30:10.20]$, and the preferred filters are
$\Ap = [5.06:6.50]$ and $\Bp = [6.50:9.50]$.
 
\subsection{Power law spectrum source}

\begin{figure}
\plotfiddle{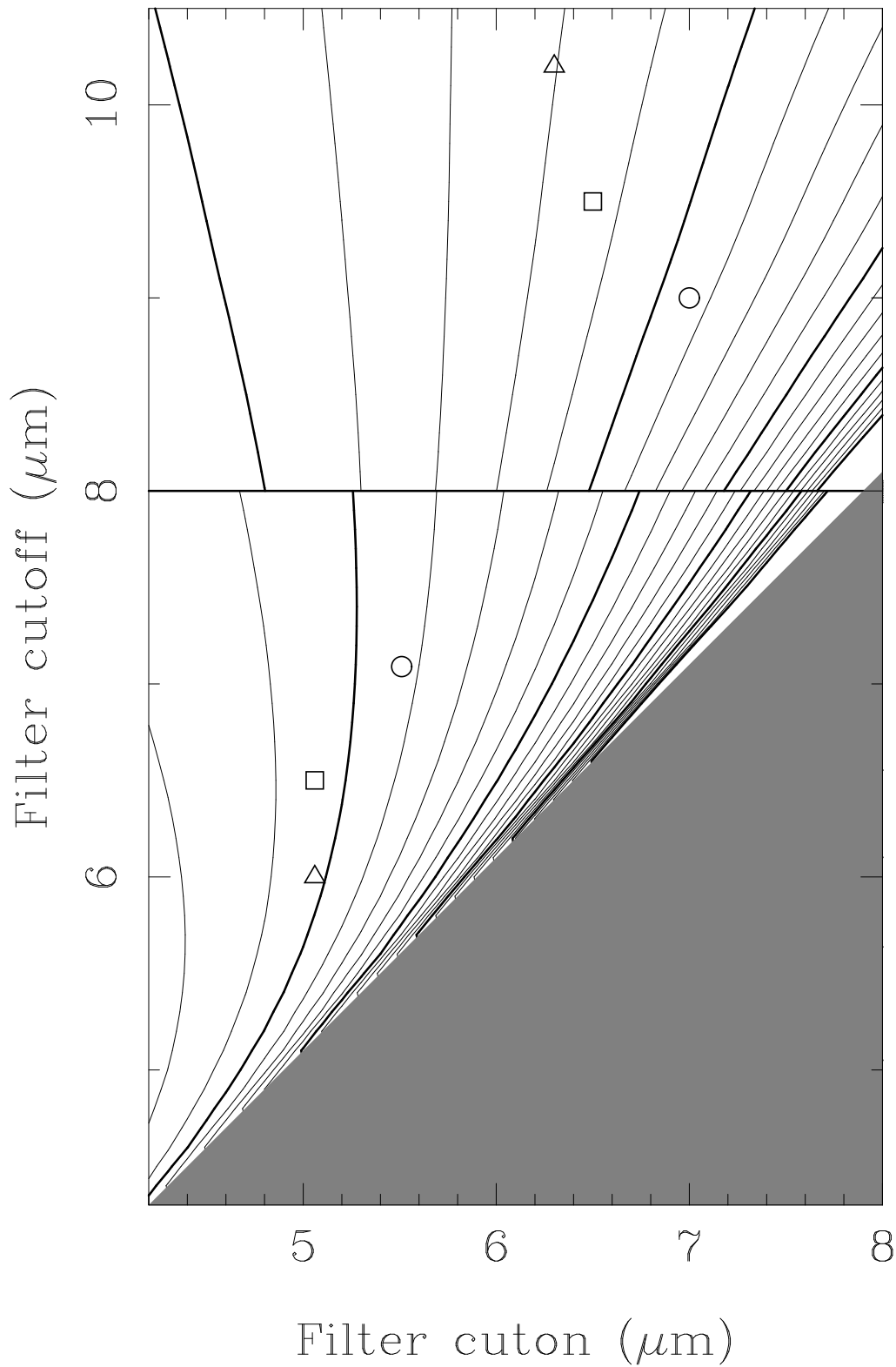}{590pt}{0}{90}{90}{-280}{-20}
\caption[]{Contours of limiting flux in $\mu$Jy ($5\sigma$, 10\,hours) as
a function of filter cuton and cutoff wavelength for a source with a
flat spectrum (in $S_\nu$). The contour range is from 0.6 to
5.0\,$\mu$Jy in steps of 0.2\,$\mu$Jy, with heavier lines at
1,2,3,4,5\,$\mu$Jy. The discontinuity at $\lambda_{\rm cutoff} =
8\,\micron$ is due to the change in coating from SiO to ZnS. Squares
mark the preferred filter specifications, triangles the optimum
filters, and circles the baseline filters.}
\label{fig:limflux}
\end{figure}

Figure~\ref{fig:limflux} is a contour plot of IRAC's background limited
$5\sigma$ sensitivity in 10\,hours of integration as a function of filter
cuton and cutoff wavelength for a source which has a flat spectrum in
$S_\nu$. Since the contour lines are fairly vertical, it is evident that
the cuton wavelength is more important than the cutoff. It is also clear
from this figure that the best sensitivity is achieved for a very blue
filter whose short wavelength cuton will be constrained by the dichroic
response. Within the parameter space plotted, a $[4.2:5.7]$ filter is most
sensitive, with a $5\sigma$ 10\,hr limiting flux of $1.1\,\mu$Jy. However,
this filter overlaps the InSb $M$ filter almost entirely, and since InSb
has a higher quantum efficiency there is little point in duplicating
it. We therefore restrict our analysis to filters with $\lambda_{\rm
cuton} > 5.06$\,\micron\ (the cutoff of the $M$ filter). The second column
in Table~\ref{tab:snr} lists the sensitivity achieved by various IRAC
filters for a flat spectrum source. The most sensitive Si:As filter is
$[5.06:7.4]$ but, as remarked above, the sensitivity is largely independent
of the cutoff wavelength, dropping only 1\% if we reduce the cutoff to
7\,\micron\ or raise it to 8.3\,\micron.

\begin{figure}
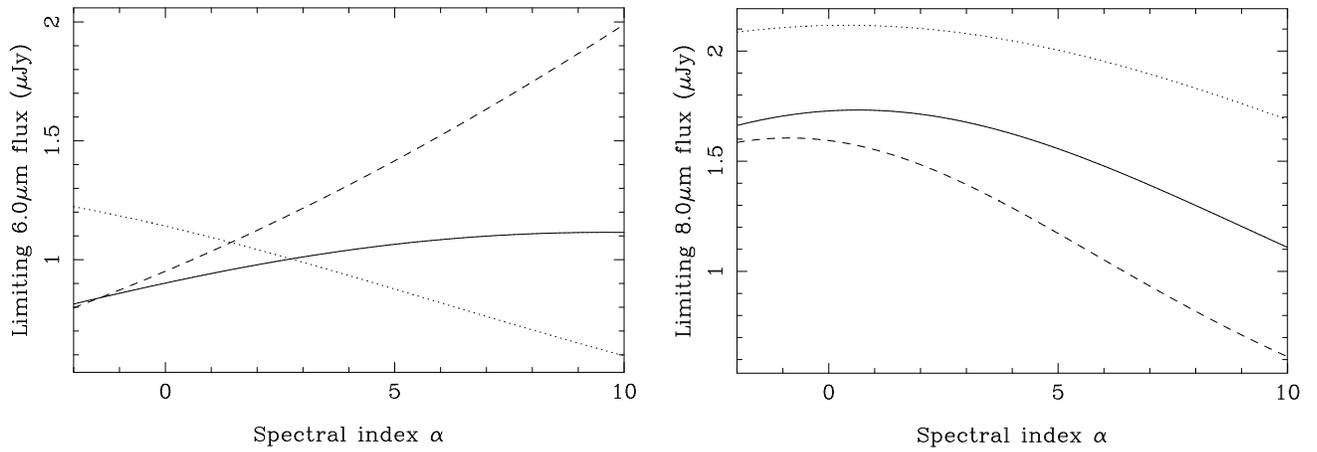

\plotfiddle{fg3a.ps}{0pt}{-90}{36}{36}{-250}{0}
\plotfiddle{fg3b.ps}{180pt}{-90}{36}{36}{0}{196}
\caption[]{Limiting sensitivity ($5\sigma$, 10\,hours) for the preferred
(solid line), optimum (dashed line), and baseline (dotted line) filters
for power laws of the form $S_\nu \propto \nu^{-\alpha}$. Left: Filter
$A$. Right: Filter $B$.}
\label{fig:plaw}
\end{figure}

The optimum and preferred filters have a $\sim 20$\% gain in sensitivity
over the baseline specifications for a flat spectrum source.
Figure~\ref{fig:plaw} shows that the improvement holds for a fairly large
range of spectral shapes, with the preferred $\Ap$ filter only losing out
to the baseline $\Ab$ filter for spectra redder than $\alpha \gtrsim 3$
($S_\nu \propto \nu^{-\alpha}$), due to the longer wavelength of the $\Ab$
filter.

\subsection{High redshift galaxies}

\begin{figure}
\plotfiddle{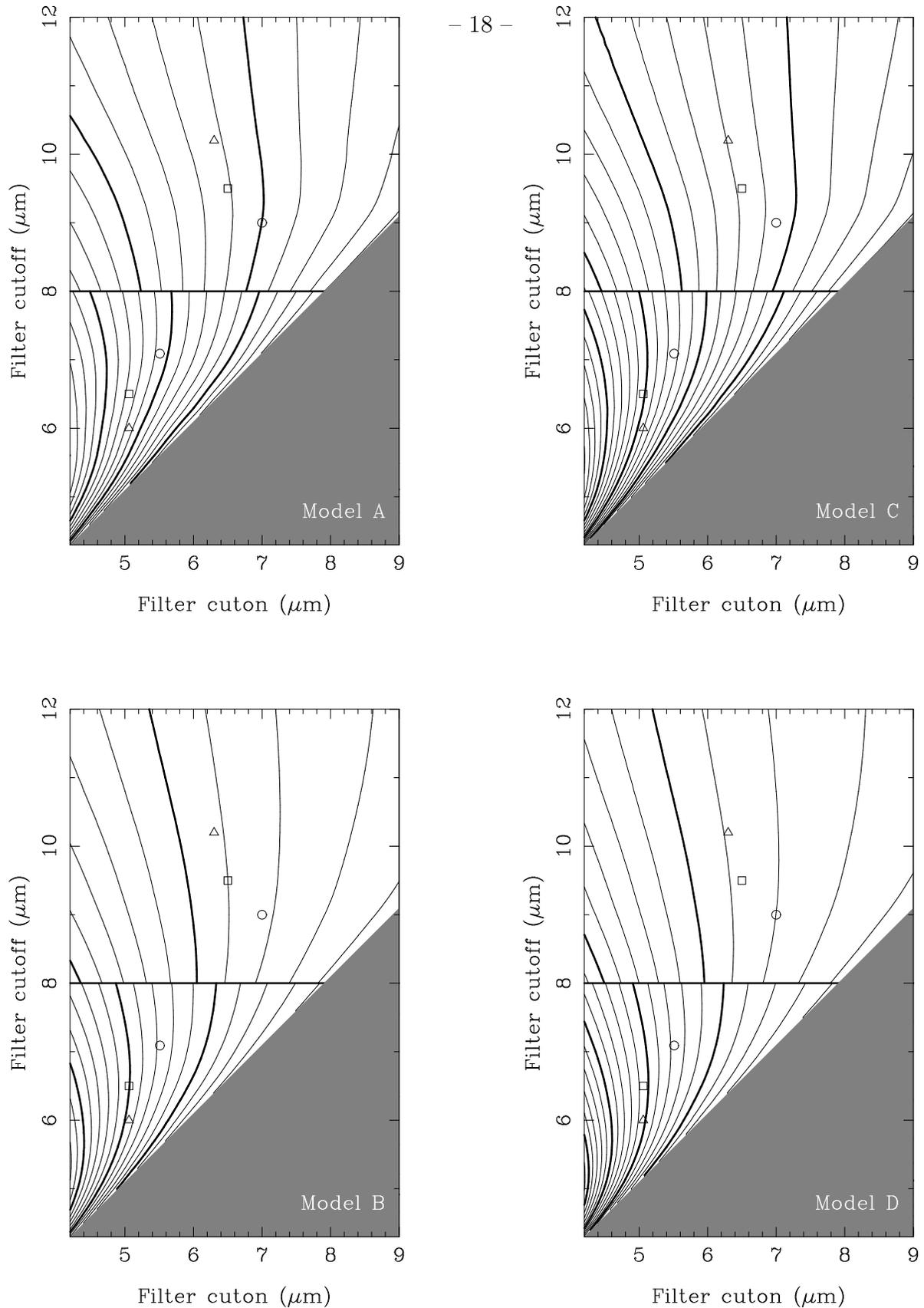}{590pt}{0}{90}{90}{-270}{-20}
\caption[]{Contours of signal-to-noise ratio as a function of filter cuton
and cutoff in a 10\,hour integration on the four galaxy models at $z=3$.
Symbols mark the filter specifications as in Figure~\ref{fig:limflux}.
Model A is contoured from 2 to 38 in steps of 2; Model B from 2 to 34 in
steps of 2; Model C from 20 to 480 in steps of 20; and Model D from 10 to
210 in steps of 10. Heavy contours are at multiples of 10,10,100,50,
respectively.}
\label{fig:snr3}
\end{figure}

Figure~\ref{fig:snr3} plots the SNR achievable in 10 hours through Si:As
filters for the four galaxy models at $z = 3$. Again the contours are
fairly vertical showing the cuton wavelength is important and the SNR
achievable is very insensitive to the cutoff. The best SNR at $z = 3$ for
a Si:As filter is again with the bluest wavelength cuton allowed
(5.06\,\micron) and with a cutoff of 7.4, 6.1, 7.2, and 6.7\,\micron\ for
models A, B, C, and D, respectively. Table~\ref{tab:snr} lists the
signal-to-noise ratios achievable at $z = 3$ for the various Si:As filter
choices.

\begin{figure}
\plotfiddle{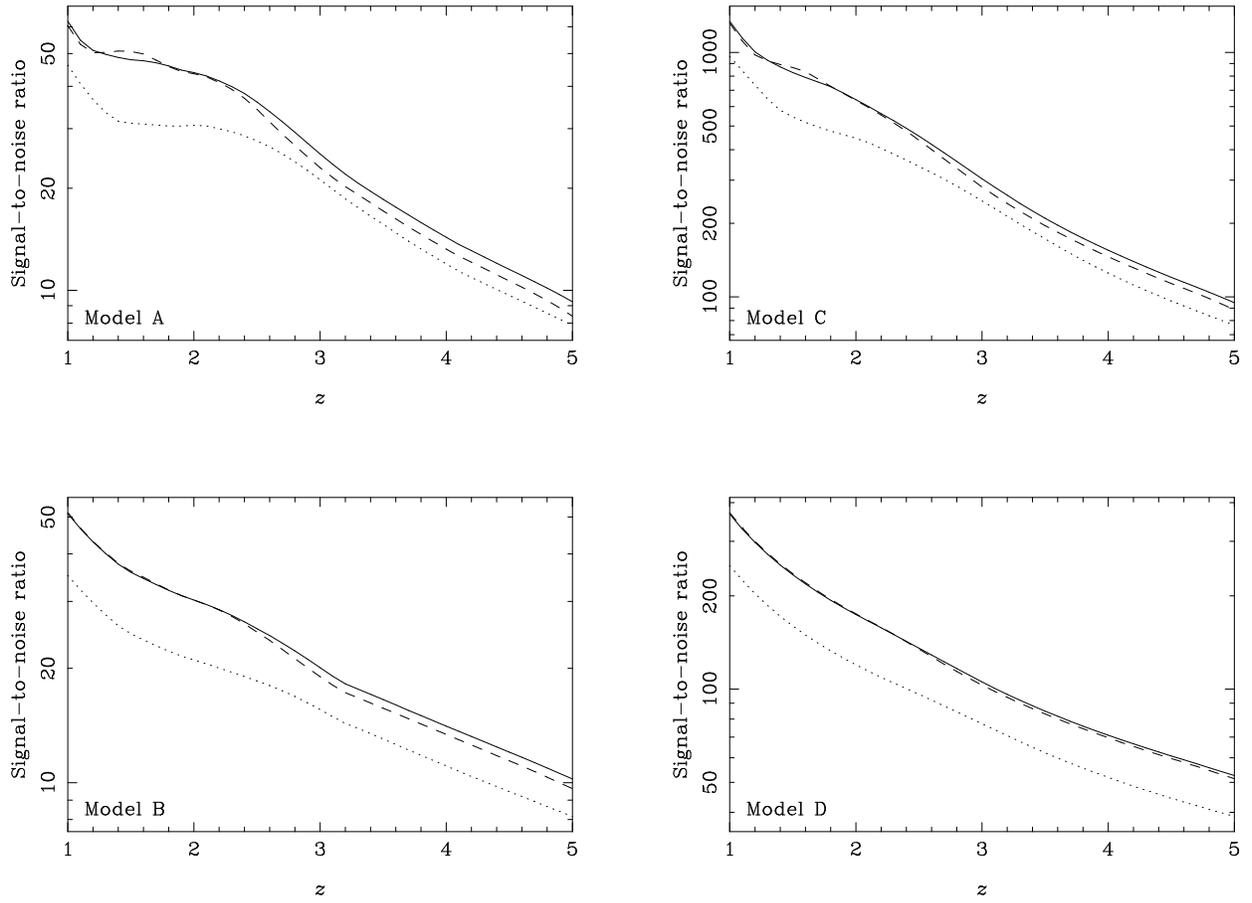}{360pt}{-90}{66}{66}{-250}{380}
\caption[]{Signal-to-noise ratio (in 10\,hours) for the three different
$A$ filters, as a function of redshift for the four galaxy models. Solid
line: preferred ($\Ap$); dashed line: optimum ($\Ao$); dotted line:
baseline ($\Ab$).}
\label{fig:snr_a}
\end{figure}

\begin{figure}
\plotfiddle{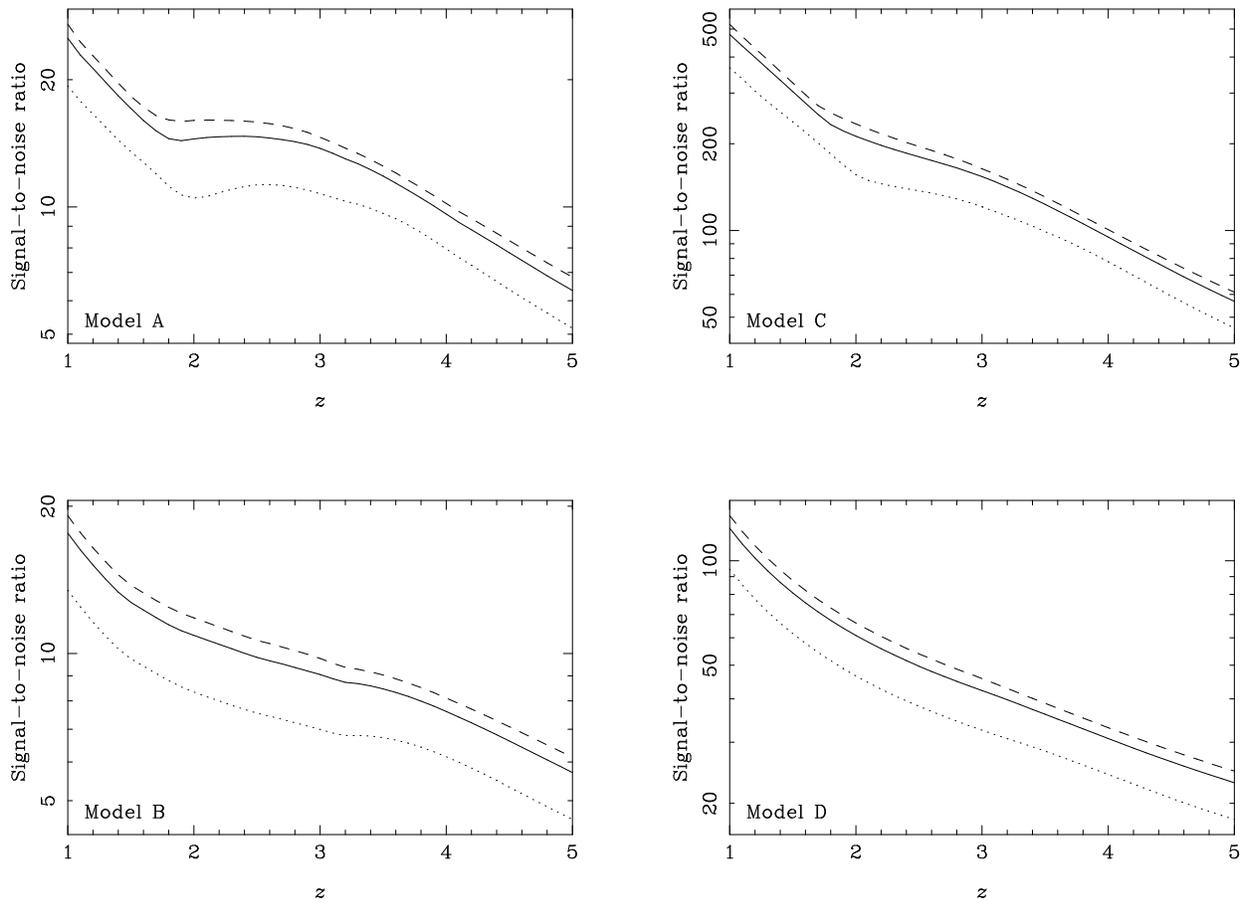}{360pt}{-90}{66}{66}{-250}{380}
\caption[]{As Figure~\ref{fig:snr_a}, but for the three $B$ filters.}
\label{fig:snr_b}
\end{figure}

\begin{figure}
\plotfiddle{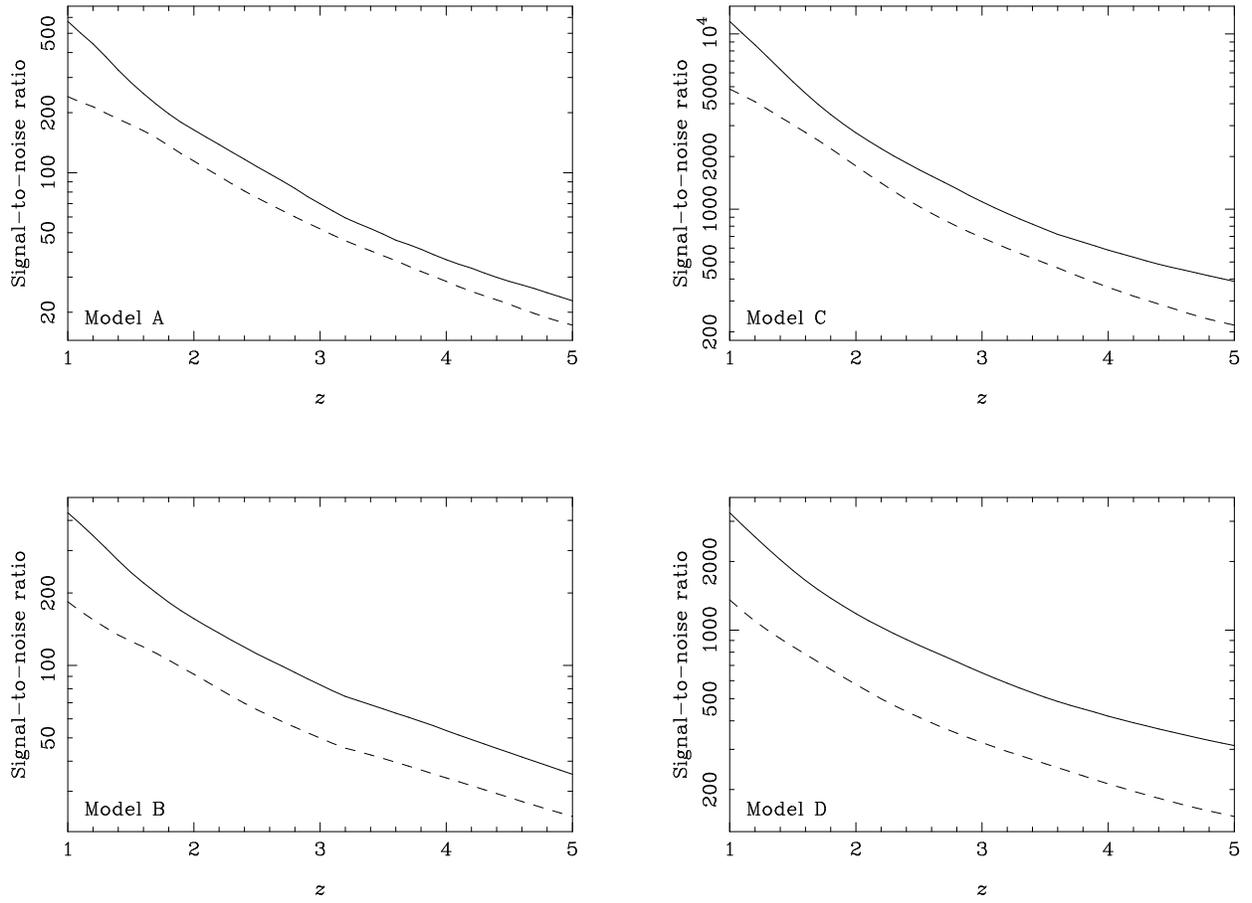}{360pt}{-90}{66}{66}{-250}{380}
\caption[]{As Figure~\ref{fig:snr_a}, but for the InSb $L$ (solid line)
and $M$ (dashed line) filters.}
\label{fig:snr_insb}
\end{figure}

The general conclusions thus far are that substantial improvements in
sensitivity can be achieved over the baseline filter specifications by
using a bluer $A$ filter and a broader $B$ filter. We show that this
applies over a broad range of redshift in Figures~\ref{fig:snr_a} and
\ref{fig:snr_b}, which plot the signal-to-noise ratio achievable in
10\,hours for the three different filter pairs. For completeness, we show
the SNR achievable through the two InSb filters as a function of redshift
in Figure~\ref{fig:snr_insb}. The higher quantum efficiency of InSb and
the lower background at short wavelengths produce an order of magnitude
increase over what is achievable with the two Si:As arrays, but the need
for the Si:As arrays will become clear in \S4.

\subsection{The {\it SIRTF\/} advantage}

The advantage in using {\it SIRTF\/} to obtain an effectively
mass-limited sample of high-redshift galaxies has already been stated.
Here, we compare the sensitivities calculated above to the limits
obtained, or obtainable, by other methods.

\begin{figure}
\plotfiddle{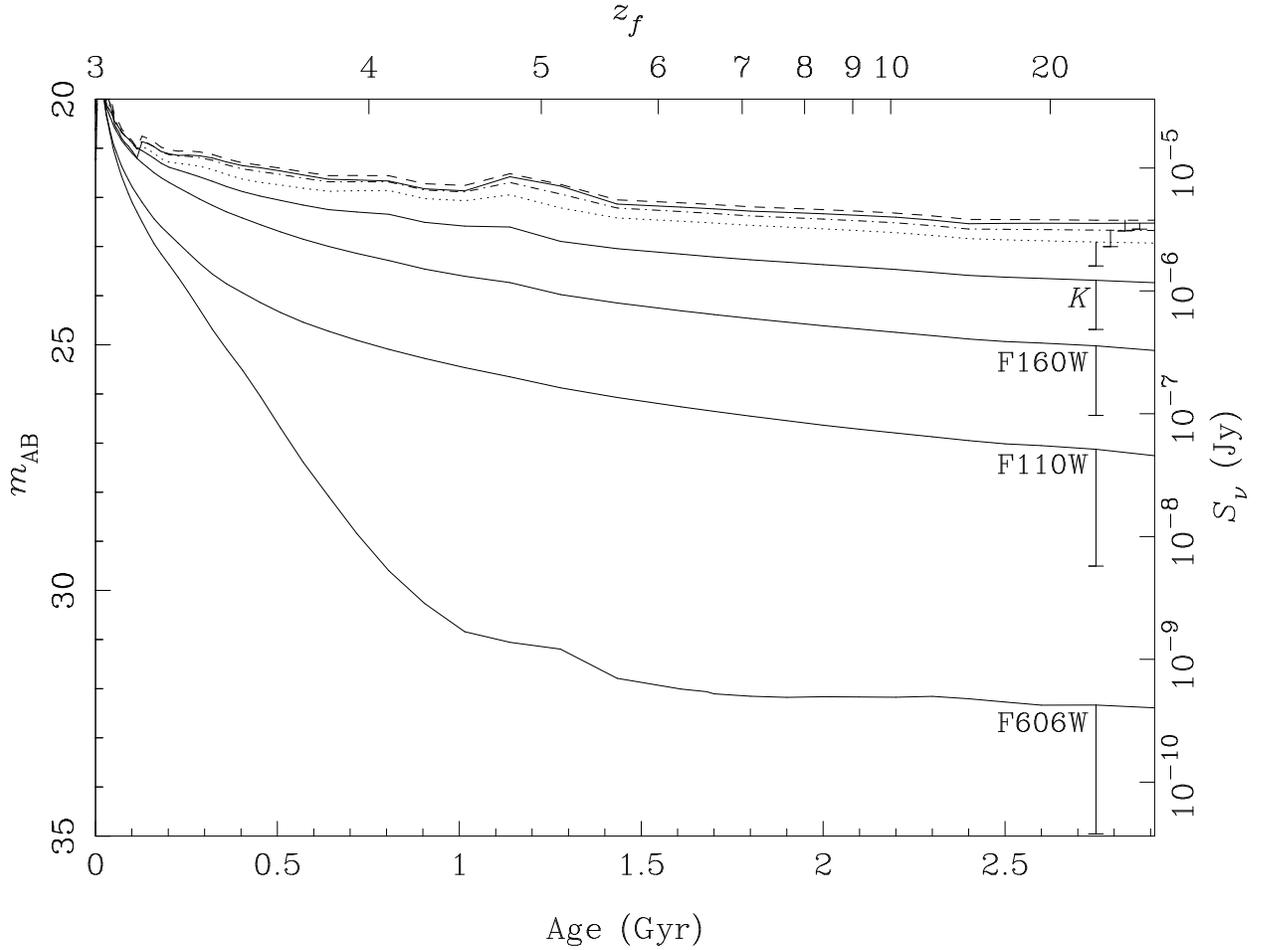}{360pt}{-90}{66}{66}{-250}{380}
\caption[]{Observed flux for an $L^*$ galaxy at $z=3$ as a function of
age (or formation redshift, $z_f$), through the filters {\it
HST\/}/WFPC2 F606W, {\it HST\/}/NICMOS F110W, {\it HST\/}/NICMOS
F160W, and ground-based $K$ (all as labeled), and the four {\it
SIRTF\/}/IRAC filters $L$ (dotted line), $M$ (dot-dash line), $\Ap$
(dashed line), and $\Bp$ (solid line). The sensitivities of the
filters are listed in Table~\ref{tab:sens} and the one-sided error
bars indicate the amount of dimming caused by $A_V = 1$.}
\label{fig:mags}
\end{figure}

In Table~\ref{tab:sens}, we first list the limits obtained, or
obtainable, by deep exposures in the optical--infrared region. In
Figure~\ref{fig:mags}, we present the observed flux of a solar
metallicity $z = 3$ $L^*$ galaxy, which was formed in an instantaneous
burst, through three different {\it HST\/} filters and our preferred
IRAC filters, as a function of age. These fluxes correspond to model A
where they intersect the right axis. From this, it is clear that
optical selection methods, such as the ``UV dropout'' technique, can
only find galaxies which have undergone substantial star formation
since $z = 4$ (i.e., similar to models C and D), as passive evolution
since then will cause their UV flux to drop below a plausible
detection threshold. Near-infrared observations fare rather better,
although it should be noted that the NICMOS HDF catalogue suffers
significant incompleteness even at high signal-to-noise ratios (the
80\% completeness limit is $m_{\rm AB} \approx 26.7$ in both filters;
Thompson et al.\ 1998). This F160W limit is sensitive enough to detect
old, sub-$L^*$ galaxies, and is comparable to the IRAC filters in this
respect. However, the F110W filter, by virtue of its sampling a region
below the 4000\,\AA\ break, is far less sensitive, and obviously there
is limited science which can be done from a single detection in one
filter. In addition, the short rest-frame wavelengths which these
filters probe are strongly affected by dust obscuration. IRAC will
therefore be the first instrument capable of detecting distant,
evolved $L^*$ (and possibly even sub-$L^*$) galaxies, if they exist,
at more than one wavelength. As a comparison, we note that the deepest
observations obtained by ISOCAM with the very wide LW-2 filter
(5.0--8.5\,\micron) produced a much higher $5\sigma$ limit of
approximately 30\,$\mu$Jy in 6 hours of integration (13 hours of
observation; Taniguchi et al.\ 1997).

\section{Photometric redshift determination at $z \sim 3$}

In this section we present the analysis which defines the ``optimum''
filters and leads to our specifications for the ``preferred'' filters.

Two factors are important in deciding upon the best filters for
photometric redshift determination. First, it is necessary to achieve a
high signal-to-noise ratio in both filters, to allow the color to be
determined accurately. This was discussed in the previous section.
Secondly, this color must be a strong function of redshift, since the
uncertainty in redshift, $\Delta z$, is related to the uncertainty in the
color, $\Delta \Gamma$ (the color defined in the sense $\Gamma \equiv
A-B$) by
\begin{equation}
\Delta z = ({\rm d}\Gamma/{\rm d}z)^{-1} \Delta \Gamma \, .
\label{eqn:dz}
\end{equation}

\begin{figure}
\plotfiddle{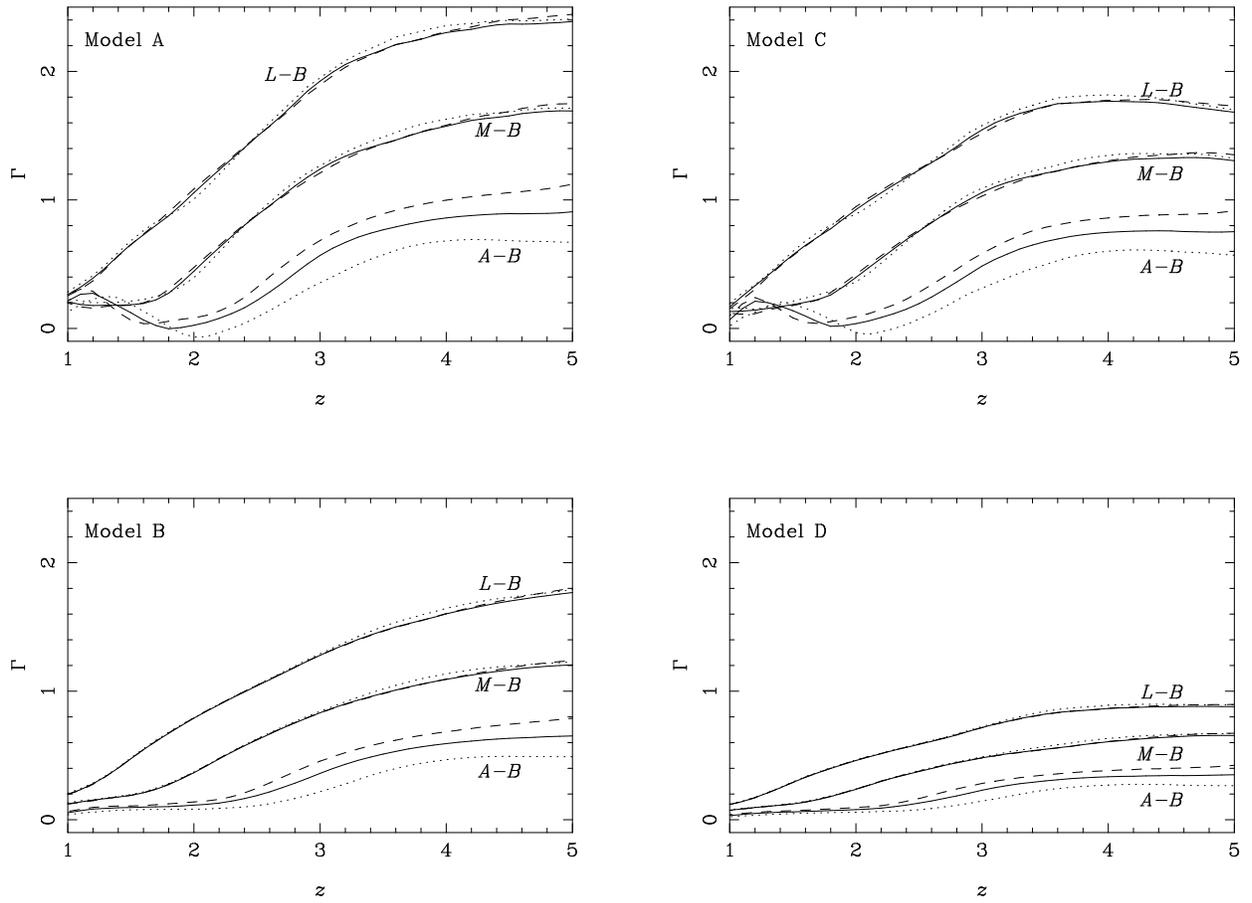}{360pt}{-90}{66}{66}{-250}{380}
\caption[]{Colours as a function of redshift for the four galaxy
models. Solid lines: preferred filters; dashed lines: optimum filters;
dotted lines: baseline filters.}
\label{fig:colz}
\end{figure}

Figure~\ref{fig:colz} plots the $L-B$, $M-B$, and $A-B$ colors as a
function of redshift for the four models. Magnitudes in each of the
filters have been determined with respect to Vega (which we model as a
9400\,K blackbody with a 2.2\,\micron\ flux of 657\,Jy). It should be
noted that at $z \lesssim 2$, the 3.3\,\micron\ PAH feature may affect
the results since it will lie in the bandpass of filter $B$. It is
apparent that the $L-M$ color is sensitive to redshift in the range $1
< z < 2$ (see the divergence between the $L-B$ and $M-B$ loci over
this redshift range), due to the 1.6\,\micron\ bump moving through the
filters. At higher redshifts the Si:As filters become increasingly
important for measuring photometric redshifts. At $z > 2$, the
increased separation between the preferred and optimum $A$ and $B$
filters, compared to the baseline filters, causes them to produce
larger color variations.  Coupled with the increased sensitivity that
these filters provide, it is clear that they will be able to produce
more accurate photometric redshifts. We now quantify the level of this
improvement.

The goal of photometric redshift determination is to minimize the
uncertainty in the derived redshift, given by Equation~\ref{eqn:dz}. We
make a few approximations to ease the calculation of this function. First,
we write ${\rm d}\Gamma/{\rm d}z \approx \Gamma(3.5) - \Gamma(2.5)$, and
make use of the fact that for well-detected ($\gtrsim 5\sigma$) sources,
the uncertainty in the measured magnitude $\Delta m \approx ({\rm
SNR})^{-1}$, where SNR is the signal-to-noise ratio. We approximate the
SNR (which varies slightly with redshift) as the geometric mean of the
actual SNR at $z = 2.5$ and $z = 3.5$. In light of these rather crude
approximations, we do not claim to be deriving accurate values of $\Delta
z$, but instead present its reciprocal, which we call $f$ (by plotting the
reciprocal, we also avoid the infinities which arise when the color change
between $z = 2.5$ and $z = 3.5$ is zero). For any two filters, $X$ and
$Y$, $f$ is therefore defined as
\begin{equation}
f(X,Y) = \frac{\Gamma(3.5) - \Gamma(2.5)}{\left(
\frac{b_X}{c_X(3.5) c_X(2.5)} +
\frac{b_Y}{c_Y(3.5) c_Y(2.5)} \right)^{1/2} },
\end{equation}
where $b_i$ is the total number of background counts and $c_i(z)$ the
total number of object counts in filter $i$ for a galaxy at redshift $z$.
Large values of $f$ are desired, and for photometric redshifts accurate to
10\% at $z \sim 3$ ($\Delta z \lesssim 0.3$), $f \gtrsim 3$ is necessary.

\begin{figure}
\plotfiddle{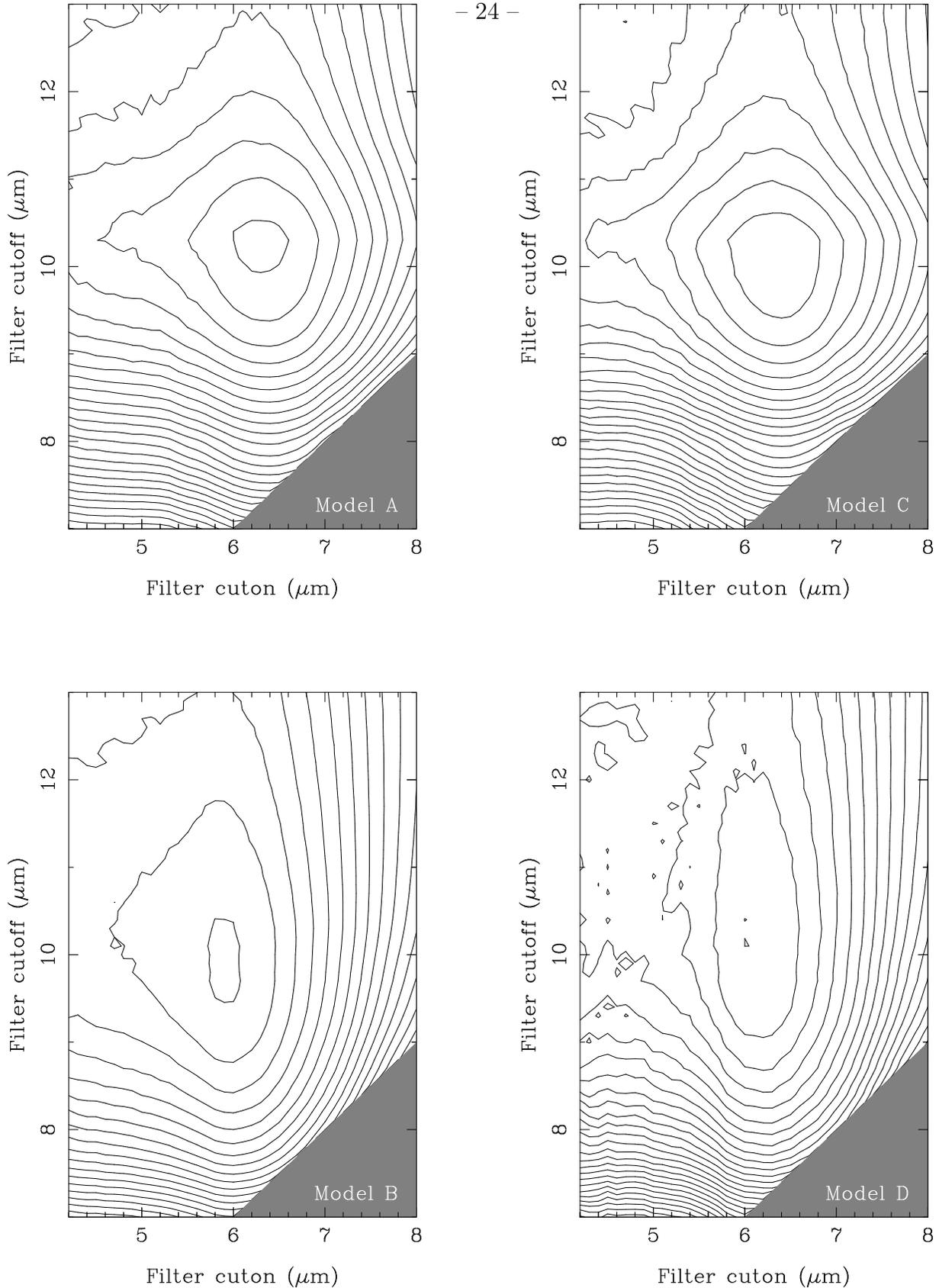}{590pt}{0}{90}{90}{-270}{-20}
\caption[]{Contours of $f$ (as defined in the text) for the InSb $M$
filter and an arbitrary (ZnS coated) Si:As filter as defined by the
axes of the plot, for each of the four galaxy models. Model A is
contoured from $-0.4$ to 6.2 in steps of 0.2; Model B from 0.0 to 3.1
in steps of 0.1; Model C from $-16$ to 58 in steps of 2; and Model D
from $-1.2$ to 6.6 in steps of 0.2.}
\label{fig:cont_m}
\end{figure}

We wish to find the specifications for the two Si:As filters that
maximize $f(A,B)$. Unfortunately, since each filter needs two parameters
to specify it, it is impossible to graphically present the results of this
four-parameter optimization in a simple form. We therefore first consider
the value of $f$ for a single Si:As filter and the InSb $M$ filter (the
results are broadly similar if the $L$ filter is used instead of $M$).
Figure~\ref{fig:cont_m} is a contour plot of $f$ for the $M$ filter and an
arbitrary Si:As filter whose specifications are given by the two axes.
This figure favors a Si:As filter centered at $\sim 8$\,\micron\ with
$\sim 50$\% bandwidth. Note that the contours are predominantly horizontal
here, so that contrary to the SNR case, the long wavelength cutoff is the
most important parameter. A broad, long wavelength filter is optimum for
determining photometric redshifts at $z \sim 3$ even though the SNR it
produces is not maximal.

We find the solution to the four-parameter optimization by performing a
grid search throughout the available parameter space. We find that,
irrespective of the particular galaxy model, the $A$ filter always has the
bluest cuton allowed (5.06\,\micron). This is to be expected, since a blue
$A$ filter produces both high signal-to-noise (Figures~\ref{fig:limflux}
and \ref{fig:snr3}) and a large color change over the redshift range of
interest. The other parameters vary slightly between the four galaxy
models, but an average solution is $A = [5.06:6.0]$ and $B = [6.3:10.2]$.
This is the how we define the ``optimum'' filter choice referred to
throughout this paper. The optimum $B$ filter has a similar specification
regardless of whether the color is formed using $L$, $M$, or $A$ as the
second filter.

Table~\ref{tab:f} lists the values of $f$ achieved for various pairs of
filters. In terms of the value of this function, the optimum filters can
produce an increase of as much as 60\% in the accuracy of the photometric
redshifts derived from the $A$ and $B$ filters, and a more modest (but
still substantial) 20\% improvement in accuracy when one of the InSb
filters is used in conjunction with the $B$ filter.

\begin{figure}
\plotfiddle{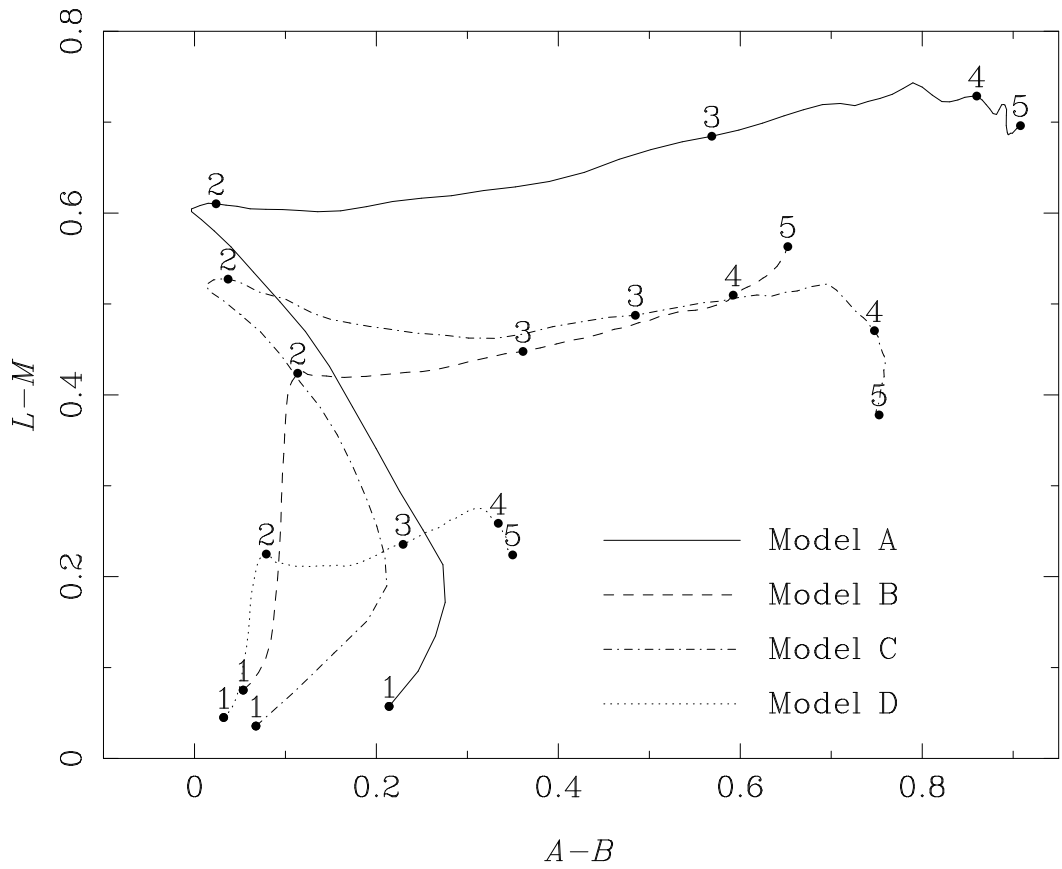}{360pt}{-90}{66}{66}{-250}{380}
\caption[]{Loci of the four galaxy models discussed in the text in the
$L-M$ {\em vs\/} $\Ap-\Bp$ color-color diagram. The locations of the
models at $z = 1$,2,3,4,5 are indicated.}
\label{fig:colcol}
\end{figure}

As an illustration of how photometric redshifts might be derived from IRAC
data, in Figure~\ref{fig:colcol} we present a $L-M$ {\em vs\/} $\Ap-\Bp$
color-color plot for the four galaxy models in the redshift range $1 < z <
5$. It can be seen that the color in the two Si:As filters is generally
able to provide an excellent measurement of the galaxy redshift for $z
\gtrsim 2$. For $1 \lesssim z \lesssim 2$, the $L - M$ color provides
most of the photometric redshift signal. Only Model D has a very limited
range of colors which might hamper analysis, since we are observing
the Rayleigh--Jeans tail of a recent starburst with only weak metal
line blanketing; however, this is exactly the sort of UV-bright galaxy
which would be detected in surveys for UV dropouts, and so this does
not pose a problem.

\section{Other considerations}

Finally, we consider the impact of filter changes on two important aspects
of distant galaxy detection which we have thus far ignored: confusion
noise, which will ultimately limit {\it SIRTF\/}'s sensitivity, and dust,
which is not included in the Bruzual \& Charlot spectral synthesis models.

\subsection{Confusion noise}

Confusion noise is expected to limit IRAC sensitivity to $\sim
0.5\,\mu$Jy ($1\sigma$; Franceschini et al.\ 1991), although this
number is quite uncertain and model-dependent. The confusion noise
also depends on the source density and the beam size, both of which
are wavelength dependent. Of particular concern is how much the
density of confusing sources will increase over the baseline filter
with a new $B$ filter, due to a longer cutoff wavelength. The longer
cutoff may significantly increase confusion because it includes the
strong, broad 7.7\,\micron\ PAH feature (e.g.\ Uchida, Sellgren \&
Werner 1998) out to $z \sim 0.5$, a large enough distance that
substantial volume is sampled and evolving dwarf galaxies may dominate
the counts.

Franceschini et al.\ (1991) estimate the density of sources at
6.7\,\micron, and their prediction of $\sim 1.2$\,arcmin$^{-2}$ above a
flux level of 30\,$\mu$Jy is consistent with Taniguchi et al.'s (1997)
detection of 15 such sources in a $3' \times 3'$ field. At microJansky
flux levels, most sources will be at $z \gtrsim 1$, so we use the
spectrum $S_\nu \propto \nu^{-0.3}$ displayed at rest wavelengths
$\sim 3$\,\micron\ to convert a limiting 8\,\micron\ flux to a
6.7\,\micron\ flux. Our best estimate of the total number of sources
above $0.5\,\mu$Jy at 8\,\micron\ is therefore $\sim
100$\,arcmin$^{-2}$, which is $\sim 0.8$ sources per 20\,pixel
detection area. We estimate the number of PAH-emitting galaxies from
the cirrus/photodissociation region spectrum and local 15\,\micron\
luminosity function of Xu et al.\ (1998). Even with very strong
($(1+z)^4$) evolution, the number of sources with $z < 0.5$ above
0.5\,$\mu$Jy is no more than $10$\,arcmin$^{-2}$, a small fraction of
the total confusing source density.

We also note that the longer cutoff wavelength of the new $\Bp$ filter
does not produce a significant increase in the diffraction-limited
beam size for a range of spectral slopes, compared to the $\Bb$ filter
(Gautier, private communication). The confusion noise will therefore
not be increased and the overall performance of the filter will not be
affected.

Finally, we note that the confusion noise limit of 0.5\,$\mu$Jy would
limit the ultimate values obtainable for $f(\Ap,\Bp)$ (as defined in
Equation~2) to 3.2, 1.5, 34, and 3.7 for Models A, B, C, and D,
respectively. In the confusion-limited case, where the photometric
accuracy is largely independent of the filter choice, it can be argued
that one should attempt to maximize the product of the photometric
redshift accuracy and the sky area which can be surveyed in a given
time. This produces slightly different specifications for the optimum
$B$ filter ([6.0:10.1] instead of [6.3:10.2]) although the arguments
of \S6 lead to the same final choice of `preferred' filters.

\subsection{Dusty galaxies}

\begin{figure}
\plotfiddle{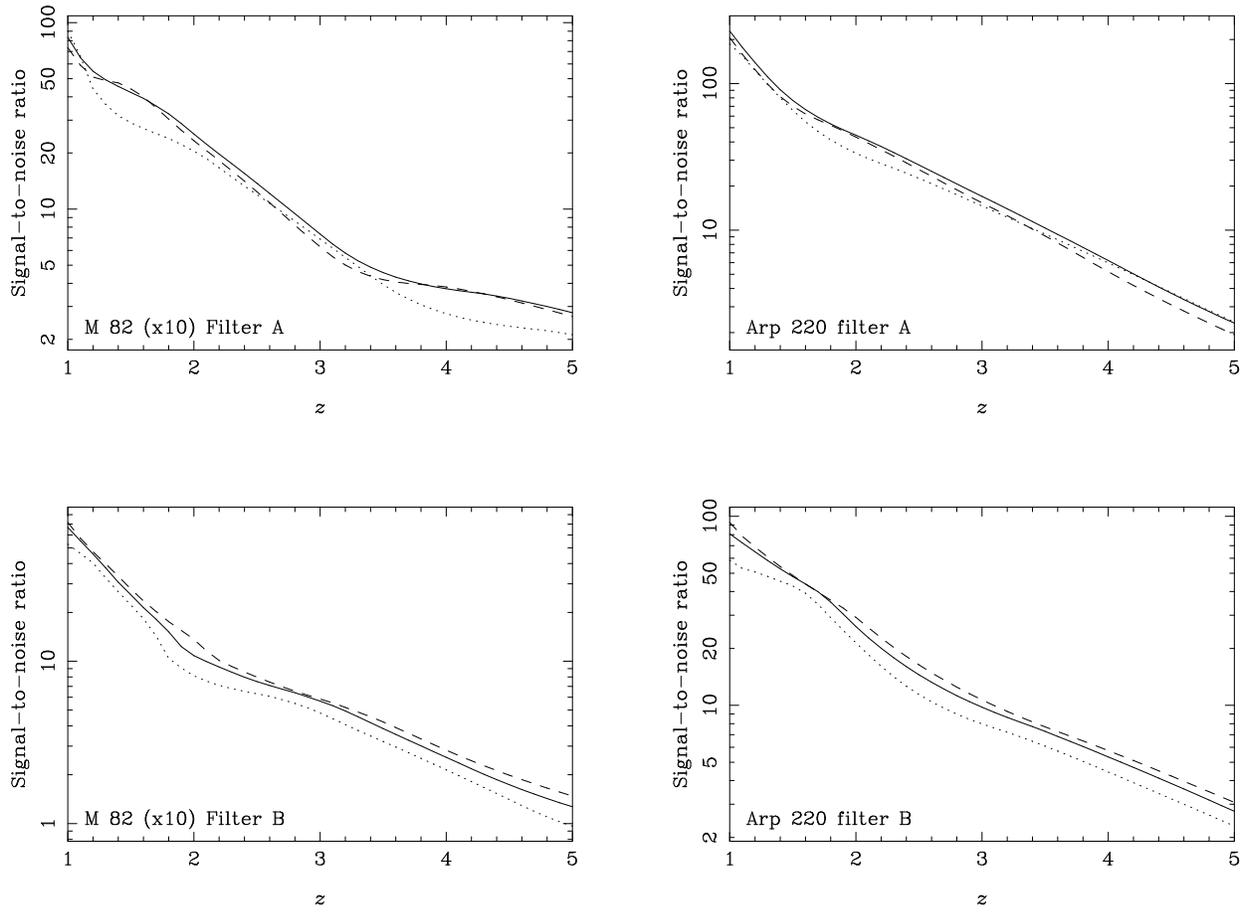}{360pt}{-90}{66}{66}{-250}{380}
\caption[]{Signal-to-noise ratio (in 10\,hours) for redshifted versions of
the starburst galaxies M~82 and Arp~220. M~82 has been scaled up by a
factor of 10 in luminosity. Solid lines: preferred filters; dashed lines:
optimum filters; dotted lines: baseline filters.}
\label{fig:m82_arp220}
\end{figure}

Nearby starburst galaxies, e.g.\ M~82 and Arp~220, are very dusty and
have red near-infrared spectra, as well as strong PAH emission
features. Our choice of a bluer $A$ filter might therefore impair our
ability to detect similar objects at high redshift, which would be
cosmologically very interesting. Figure~\ref{fig:plaw} shows that for
spectra redder than $\alpha \gtrsim 3$, the $\Ab$ filter is more
sensitive than $\Ap$. In Figure~\ref{fig:m82_arp220} we show the
sensitivities of the three different filter pairs to these two
starburst galaxies as a function of redshift (the spectrum of M~82 has
been scaled up by a factor of 10 in luminosity in this figure). The
spectral energy distributions required to make these figures were
produced with photometric data culled from a variety of sources, and
aperture matching using curve-of-growth analysis was required. This is
likely to produce errors where the spectral region sampled by a given
filter includes the transition between different datasets (e.g.\
optical and near-infrared), and the signal-to-noise ratios (both
absolute and relative) should probably not be trusted to more than
$\sim 10$\%. Given these caveats, there is little to choose between
the different $A$ filters, but the sensitivity improvement offered by
the $\Bp$ filter is unambiguous.

\section{Conclusions: the ``preferred'' filters}

Although the broad $\Bo$ filter is supported by sensitivity calculations
and does not suffer significantly from increased confusion noise, a
somewhat narrower filter makes sense for other reasons. The almost 50\%
bandwidth of the $\Bo$ filter will make photometry difficult and may
compromise other uses of this filter. In particular, the strong silicate
absorption feature at 9.7\,\micron\ falls in the $\Bo$ bandpass. We
therefore recommend a shorter cutoff wavelength, and propose $B =
[6.5:9.5]$ (a 37\% bandpass) as our ``preferred'' $B$ filter, $\Bp$.

Conversely, the optimum $\Ap = [5.06:6.0]$ choice is only a 17\%
bandwidth, and does not produce optimum SNR (see Table~\ref{tab:snr}). It
does however avoid the PAH feature at 6.2\,\micron, offering the
possibility of using filters $A$ and $B$ as a PAH diagnostic. However the
$M$ filter also acts as a good continuum filter for this problem, with
better SNR and the advantage that it shares the same field as the $B$
filter. Hence we recommend a broader $A$ filter filling the wavelength
regime between $M$ and $B$. Our preferred $A$ filter, $\Ap$, is then
$[5.06:6.5]$, with a 25\% bandpass.

These preferred filters reduce the value of $f$ by about 10\%, compared to
the optimum filters, but we feel that this is outweighed by the gain in
general scientific usefulness. In any case, they still offer a significant
improvement over the baseline 6.3 and 8.0 filters. These specifications
have now been adopted for the IRAC instrument.

\section*{Acknowledgments}

The authors wish to thank Roc Cutri, Nick Gautier, Craig McCreight, and
Harvey Moseley for their help. This work was performed at the Jet
Propulsion Laboratory, California Institute of Technology, under a
contract with the National Aeronautics and Space Administration.

\clearpage

\clearpage

\begin{table}
\caption[]{Signal-to-noise ratios achievable with IRAC in 10 hours.}
\label{tab:snr}
\begin{center}
\begin{tabular}{lrrrrr}
\hline \hline
& \multicolumn{5}{c}{SNR for Model} \\
Filter         & 10\,$\mu$Jy & \cntr{A} & \cntr{B} & \cntr{C} & \cntr{D}
\\
\hline
Best filter\tablenotemark{a} & 55.9 & 26.9 & 20.0 & 309 & 105 \\
$\lambda_{\rm cutoff}$ & [6.9]  & [7.1] & [6.7] & [6.9] & [6.4] \\
\hline
$L$   &  232.2  & 70.9 & 82.4 & 1104 & 649 \\
$M$   &  136.7  & 53.4 & 49.5 &  688 & 321 \\
$\Ab$ &   35.4  & 21.8 & 15.4 &  247 &  77 \\
$\Ao$ &   52.6  & 23.6 & 18.8 &  281 & 103 \\
$\Ap$ &   55.5  & 25.9 & 19.9 &  304 & 105 \\
$\Bb$ &   23.6  & 10.3 &  6.4 &  112 &  30 \\
$\Bo$ &   31.4  & 12.8 &  8.3 &  139 &  39 \\
$\Bp$ &   28.9  & 12.6 &  7.9 &  136 &  37 \\
\hline
\end{tabular}
\end{center}
\tablenotetext{a}{The ``best filter'' is the most sensitive for a
10\,$\mu$Jy flat spectrum source and the four galaxy models at $z =
3$, with the constraint that no filter extends bluer than
5.06\,\micron. All such filters cut on at 5.06\,\micron\ and extend to
the cutoff wavelength listed.}
\end{table}

\clearpage

\begin{table}
\caption[]{Sensitivities ($5\sigma$, 10\,hours) of instrument/filter
combinations.}
\label{tab:sens}
\begin{center}
\begin{tabular}{lcr@{.}l}
\hline \hline
Instrument/filter & $m_{\rm AB}$ & \multicolumn{2}{c}{$S_\nu$ ($\mu$Jy)} \\
\hline
{\it HST\/} WFPC2/F606W\tablenotemark{a}  & 28.4 &  0&016 \\
{\it HST\/} NICMOS/F110W\tablenotemark{a} & 27.2 &  0&047 \\
{\it HST\/} NICMOS/F160W\tablenotemark{a} & 27.5 &  0&037 \\
Subaru IRCS/$K$\tablenotemark{b}          & 24.4 &  0&63  \\
{\it ISO\/} CAM/LW-2\tablenotemark{c}     & 20.4 & 24&7   \\
{\it SIRTF\/} IRAC/$L$   & 25.6 &  0&21  \\
{\it SIRTF\/} IRAC/$M$   & 25.0 &  0&36  \\
{\it SIRTF\/} IRAC/$\Ap$ & 24.0 &  0&90  \\
{\it SIRTF\/} IRAC/$\Bp$ & 23.3 &  1&73  \\
\hline
\end{tabular}
\end{center}
\tablenotetext{a}{Scaled from observations of the Hubble Deep Field
(Williams et al.\ 1996; Thompson et al.\ 1998)}
\tablenotetext{b}{A description of IRCS can be found in Tokunaga et
al.\ (1998)}
\tablenotetext{c}{Scaled from Taniguchi et al.'s (1997) observations of
the the Lockman Hole}
\end{table}

\clearpage

\begin{table}
\caption[]{Values of $f$ for various filter pairs.}
\label{tab:f}
\begin{center}
\begin{tabular}{ccrrrr}
\hline \hline
& & \multicolumn{4}{c}{$f$ for Model} \\
\multicolumn{2}{c}{Filters} & \cntr{A} & \cntr{B} & \cntr{C} & \cntr{D} \\
\hline
\multicolumn{6}{c}{Baseline filters} \\
\hline
$L$   & $\Bb$ & 6.4 & 3.1 & 57 & 7.6 \\
$M$   & $\Bb$ & 5.4 & 2.6 & 52 & 5.7 \\
$\Ab$ & $\Bb$ & 4.1 & 1.6 & 41 & 4.3 \\
\hline
\multicolumn{6}{c}{Optimum filters} \\
\hline
$L$   & $\Bo$ & 7.6 & 3.7 & 66 & 9.1 \\
$M$   & $\Bo$ & 6.2 & 3.1 & 59 & 6.6 \\
$\Ao$ & $\Bo$ & 6.0 & 2.8 & 63 & 6.7 \\
\hline
\multicolumn{6}{c}{Preferred filters} \\
\hline
$L$   & $\Bp$ & 7.3 & 3.5 & 65 & 8.8 \\
$M$   & $\Bp$ & 6.0 & 3.0 & 58 & 6.4 \\
$\Ap$ & $\Bp$ & 5.6 & 2.5 & 58 & 6.1 \\
\hline
$\Ao$ & $\Bp$ & 5.9 & 2.7 & 62 & 6.6 \\
\hline
\end{tabular}
\end{center}
\end{table}

\end{document}